\documentclass[final,3p,times,twocolumn]{elsarticle}

\usepackage[utf8]{inputenc}
\usepackage{amsmath,amssymb,exscale}
\usepackage{graphicx}
\usepackage{epsfig}
\usepackage{multicol}
\usepackage{color}
\usepackage{xcolor}
\usepackage{hyperref}
\usepackage{mathrsfs}
\hypersetup{colorlinks,bookmarksopen,bookmarksnumbered,citecolor=rossos,
linkcolor=black,pdfstartview=FitH,urlcolor=blus}
\usepackage{amsfonts}
\usepackage{slashed}
\usepackage{blindtext}
 \usepackage{fancyhdr}
\usepackage{mathtools}
\usepackage{cancel,xcolor}
\biboptions{numbers,sort&compress}
\usepackage{tensor}
\usepackage{soul}
\usepackage{accents}

\usepackage{xcolor,pifont}
\newcommand*\colourcheck[1]{%
  \expandafter\newcommand\csname #1check\endcsname{\textcolor{#1}{\ding{52}}}%
}
\colourcheck{blue}
\colourcheck{green}

\allowdisplaybreaks

\newcommand{\be}{\begin{equation}}
\newcommand{\ee}{\end{equation}}
\newcommand{\bear}{\begin{array}}
\newcommand{\eear}{\end{array}}
\newcommand{\ba}{\begin{eqnarray}}
\newcommand{\ea}{\end{eqnarray}}

\newcommand{\bea}{\begin{equation}\begin{aligned}} 
\newcommand{\eea}{\end{aligned}\end{equation}}

\def\b{\beta}

\def\e{\epsilon}

\def\k{\kappa}
\def\l{\lambda}
\def\m{\mu}
\def\n{\nu}

\def\p{\pi}

\def\r{\rho}
\def\s{\sigma}

\def\x{\xi}

\def\tns{\tensor}

\newcommand{\td}{{\rm d}}
 
\definecolor{blue(pigment)}{rgb}{0.2, 0.2, 0.6}
\definecolor{coolblack}{rgb}{0.0, 0.18, 0.39}
\definecolor{darkblue}{rgb}{0.0, 0.0, 0.55}
\definecolor{darkcerulean}{rgb}{0.03, 0.27, 0.49}
\definecolor{violetred}{rgb}{0.8,0.13,0.56}
\definecolor{fuxia}{rgb}{1,0,1}
\definecolor{orchid}{rgb}{0.85,0.44,0.84}
\definecolor{lightpink}{rgb}{1.00,0.71,0.76}

\newcommand{\R}[1]{{\color{red} #1}}

\begin{document}

\begin{frontmatter}

\title{\vspace{-1cm}\huge {\color{blue(pigment)}
Electroweak vacuum decay in metric-affine gravity } \vspace{0.5cm}}

 \author[]{\large {\bf Ioannis D. Gialamas}\corref{cor1}}
 \cortext[cor1]{Email address: ioannis.gialamas@kbfi.ee}
  \author[]{\large {\bf Hardi~Veerm\"ae}\corref{cor2}}
 \cortext[cor2]{Email address: hardi.veermae@cern.ch}
\address{\normalsize Laboratory of High Energy and Computational Physics, National Institute of Chemical Physics and Biophysics, \\R{\"a}vala pst.~10, 10143, Tallinn, Estonia}

\begin{abstract}
We investigate the stability of the electroweak vacuum in metric-affine gravity in which the Standard Model Higgs boson can be non-minimally coupled to both the Ricci scalar and the Holst invariant. We find that vacuum stability is improved in this framework across a wide range of model parameters.
\end{abstract}

\end{frontmatter}
 
 \begingroup
\hypersetup{linkcolor=blus}
\endgroup

\vspace{-0.1cm}
\section{Introduction}\label{introduction}
\vspace{-0.1cm}
It is well known that the potential of the Higgs boson in the Standard Model (SM) is deeper at high energies than in the electroweak vacuum permitting its decay through quantum tunneling~\cite{Coleman:1977py,Arnold:1989cb,Sher:1988mj,Arnold:1991cv}. Although this does not invalidate the SM, the electroweak vacuum is predicted to be metastable in the absence of contributions from UV physics~\cite{Sher:1993mf,Casas:1994qy,Isidori:2001bm,Espinosa:2007qp,Elias-Miro:2011sqh,Degrassi:2012ry, Buttazzo:2013uya,DiLuzio:2015iua,Chigusa:2017dux}. 

Coleman and De Luccia~\cite{Coleman:1980aw} were the first to delve into the matter of gravitational effects on vacuum decay. 
Subsequently, multiple studies of gravitational corrections have been performed~\cite{Isidori:2007vm, Branchina:2016bws, Rajantie:2016hkj, Czerwinska:2016fky, Salvio:2016mvj, Markkanen:2018pdo, Espinosa:2020qtq, Devoto:2022qen, Gialamas:2022gxv}, along with discussions about the impact of black holes on the amplification or reduction of the vacuum decay rate~\cite{Gregory:2013hja, Burda:2015isa, Burda:2015yfa, Burda:2016mou, Tetradis:2016vqb, Canko:2017ebb, Kohri:2017ybt, Gorbunov:2017fhq, Mukaida:2017bgd, Gregory:2018bdt, Hayashi:2020ocn,  Shkerin:2021zbf, Shkerin:2021rhy, DeLuca:2022cus, Strumia:2022jil,  Briaud:2022few, Gregory:2023kam}.

In this letter, we extend the calculation of gravitational corrections to vacuum decay in the context of metric-affine gravity. There, in contrast to general relativity, the connection is taken to be an independent variable without the usual symmetries of the Levi-Civita one. As a result, the Riemann tensor does not possess the symmetries it has in the metric case, and thus the gravitational action should be extended by including an additional scalar curvature invariant - the Holst invariant. This term commonly appears in Loop Quantum Gravity~\cite{Rovelli:1997yv} and has been studied in various branches of high energy physics, such as black hole thermodynamics~\cite{Domagala:2004jt, Meissner:2004ju}. Lately, its significance in inflationary cosmology~\cite{Langvik:2020nrs, Shaposhnikov:2020gts, Piani:2022gon, Pradisi:2022nmh, Salvio:2022suk, Gialamas:2022xtt}, and high energy physics phenomenology~\cite{Shapiro:2014kma, Shaposhnikov:2020frq, Shaposhnikov:2020aen, Karananas:2021zkl, Karananas:2021gco}, has garnered a great deal of attention. 

\vspace{-0.1cm}
\section{The gravitational action}
\label{sec_grav:act}
\vspace{-0.1cm}

Metric-affine theories of gravity treat the metric tensor $g_{\m\n}$ and the connection $\tns{\Gamma}{^\l_\m_\n}$ as independent variables. This should be contrasted with the usual metric gravity that uses the Levi-Civita connection $\left\{\tns{}{^\l_\m_\n}\right\}$, which is completely determined by the metric.
It is useful to decompose the connection $\tns{\Gamma}{^\l_\m_\n}$ as
\be
\label{eq:Distor}
    \tns{\Gamma}{^\l_\m_\n} \equiv \left\{\tns{}{^\l_\m_\n}\right\} + \tns{C}{^\l_\m_\n} \,,
\ee
where $\tns{C}{^\l_\m_\n}$ is dubbed the distortion tensor.
The Riemann tensor $\tns{\mathcal{R}}{^\m_\n_\r_\s}$ is constructed from the connection $\Gamma$ in the usual way and one can form two scalars that are linear on it. These are the Ricci scalar and the Holst invariant~\cite{Hojman:1980kv, Holst:1995pc}
\be 
\label{eq:Ricci_Holst}
    \mathcal{R} \equiv \tns{\mathcal{R}}{^\m^\n_\m_\n}\, , 
    \qquad
    \widetilde{\mathcal{R}} \equiv \frac{1}{2}\tns{\epsilon}{^\m^\n^\r^\s}\tns{\mathcal{R}}{_\m_\n_\r_\s}\,,
\ee
where $g$ is the determinant of the metric tensor and $\epsilon^{\m\n\r\s}$ is the totally antisymmetric tensor. In metric gravity, the Holst invariant vanishes identically due to the symmetries of the Riemann tensor.

The most general action linear in the Riemann tensor and containing terms of at most dimension 4 has the form\footnote{Natural units with $c = \hbar = 1$ are used throughout this paper.}
\be
\label{eq:Action_1}
  S = \int{\td}^4 x \sqrt{-g} \bigg[ \frac{\mathcal{R}}{2} f(h) + \frac{\widetilde{\mathcal{R}}}{2} \tilde{f}(h) -  \frac{1}{2} (\partial h)^2 - V(h) \bigg]\,,
\ee
where $V(h)$ is the Higgs potential, 
\be
\label{def:fg}
    f(h) = 1/\k +\xi h^2, \qquad
    \tilde{f}(h) = \beta/\k + \tilde{\xi} h^2\,,
\ee
are non-minimal couplings, $\beta$, $\xi$ and $\tilde \xi$ are constant couplings, and $\k = 1/M_{\rm Pl}^2$. The function $1/\tilde{f}(h)$ can be thought of as a field-dependent Barbero-Immirzi parameter~\cite{Holst:1995pc, Immirzi:1996di}. We will first work out the formalism without assuming a specific functional form of $f$, $\tilde{f}$, and $V$ and suppress their arguments for notational brevity.

In addition to the Ricci and Holst terms, metric-affine gravity permits the construction of 20 additional scalars with mass dimension 2 from torsion  {
 $T_{\r\m\n} \equiv 2 \Gamma_{\r[\m\n]}$
and non-metricity $Q_{\r\m\n} \equiv \nabla_{\r}g_{\m\n}$
~\cite{Iosifidis:2021bad,Rigouzzo:2022yan}. Although such terms are not considered in this work, our perturbative results can be straightforwardly extended to include them as will be explained below. We will also neglect couplings between the connection and fermions as they will only generate Planck-suppressed four-fermion and higher-order scalar-fermion interactions~\cite{Latorre:2017uve, Shaposhnikov:2020frq, Karananas:2021zkl, Pradisi:2022nmh} and do not affect the leading order corrections to vacuum stability.

We remark that the action \eqref{eq:Action_1} also appears in Einstein-Cartan gravity.
A crucial distinction with the current case is that the Einstein-Cartan connection is decomposed using the Levi-Civita connection and torsion and, unlike in metric-affine gravity, the non-metricity is taken to be zero. Nevertheless, as we will show below, the metric-affine framework contains both the metric and Palatini formulations as limiting cases and the non-metricity may be taken to zero without loss of generality.

In order to study bounce solutions, we construct the Euclidean action~\eqref{eq:Action_1} 
by analytically continuing the Lorentzian signature $(-,+,+,+)$ to the Euclidean one $(+,+,+,+)$. 
Then, to bring the action to a more conventional form, we will first integrate out the connection. To this aim, we will express the Ricci scalar and the Holst invariant in terms of the metric Ricci scalar $R[g]$ and the distortion tensor,
\begin{subequations}
\begin{align}
\label{eq:ricci_b}
    \mathcal{R}
    =& R+D_{ \mu}\tns{C}{^\m_\n^\n}-D_{ \nu}\tns{C}{^\m_\m^\n}+\tns{C}{^\m_\m_\l}\tns{C}{^\l_\n^\n}-\tns{C}{^\m_\n_\l}\tns{C}{^\l_\m^\n}\,,
\\
\widetilde{\mathcal{R}}=& \e^{\m\n\r\s}\left(D_{\m}\tns{C}{_\r_\n_\s}+\tns{C}{_\r_\m_\l}\tns{C}{^\l_\n_\s}\right)\,,
\label{eq:Holst_b}
\end{align}
\end{subequations}
where $D$ denotes the covariant derivative of the Levi-Civita connection. Substituting Eq.~\eqref{eq:ricci_b} and~\eqref{eq:Holst_b} into the action~\eqref{eq:Action_1}, yields\footnote{Although the Holst term picks up an imaginary unit when continuing to Euclidean space, analogously to the CP violating topological term in Yang-Mills theory (e.g., see Ref.~\cite{DiLuzio:2020wdo}), its effect is negated due to the dependence of $\e^{\m\n\r\s}\e_{\m\n\r\s} = {\rm sign}(g) \, 4!$ on the sign of the metric determinant.}
\begin{align}
\label{eq:Action_2}
    S_{E} 
=& \int{\td}^4 x \sqrt{g} \bigg[ - \frac{R}{2} f +  \frac{1}{2} (\partial h)^2 + V  \nonumber
\\  & - \frac{f}{2} \left( D_{ \mu}\tns{C}{^\m_\n^\n}-D_{ \nu}\tns{C}{^\m_\m^\n}
    +   \tns{C}{^\m_\m_\l}\tns{C}{^\l_\n^\n}-\tns{C}{^\m_\n_\l}\tns{C}{^\l_\m^\n} \right) \nonumber
\\  & -  \frac{i\tilde{f}}{2}\e^{\m\n\r\s}\left(D_{ \m}\tns{C}{_\r_\n_\s}+\tns{C}{_\r_\m_\l}\tns{C}{^\l_\n_\s}\right)  \bigg]\,.
\end{align}
The distortion tensor obeys an algebraic non-homogeneous linear equation of motion. Thus, in order to integrate it out in full generality, it is sufficient to find a particular solution to this equation~\cite{Pradisi:2022nmh}. Such a solution is given by~\cite{Gialamas:2022xtt}
\be
\label{eq:Dist_sol}
    C_{\mu\nu\rho}
    \,= \frac{1}{2} \left(\,g_{\nu\mu}\partial_{\rho}X - g_{\nu\rho}\partial_{\mu}X - i \epsilon_{\mu\nu\rho\sigma}\partial^{\sigma}Y \right)\,,
\ee
where
\be
\label{eq:Dist_sol_XY}
    f = e^X\cos(Y), \qquad
    \tilde{f} = e^X\sin(Y) \, .
\ee
This solution is metric compatible, i.e., $Q_{\r\m\n} = - 2C_{(\n|\r|\m)}= 0$, and has torsion $T_{\r\n\m} = 2C_{\r[\n\m]}$. So, the theory is dynamically equivalent to the Einstein-Cartan theory. On the other hand, the particular solution \eqref{eq:Dist_sol} is not the general one because of the projective symmetry of the action, $C_{\r\n\m} \to C_{\r\n\m} + g_{\r\m}A_{\n}$, which can be used to induce the non-metricity $Q_{\r\m\n} = - 2g_{\m\n}A_{\r}$. In particular, the Palatini limit with $\tilde{f} = 0$ is obtained by choosing $A_{\n} = \partial_{\n} X/2$. 

Substituting \eqref{eq:Dist_sol} in the action~\eqref{eq:Action_2} gives
\be
\label{eq:Action_3b}
    S_E = \int{\td}^4 x \sqrt{g} \left[ - \frac{R}{2} f +  \frac{1}{2}K (\partial h)^2+ V \right]\,,
\ee
where the contribution of the independent connection is now fully captured by the kinetic function\footnote{The primes denote differentiation with respect to the argument of the function, that is, depending on the context, with respect to $h$ or $r$.}
\be
\label{eq:K}
    K = 1 + \frac{3}{2}\frac{f\tilde{f}'^2-2f'\tilde{f}\tilde{f}'-ff'^2}{f^2+\tilde{f}^2}\,.
\ee
Note that the action \eqref{eq:Action_3b} does not depend on the sign of $\tilde{f}$.

For specific combinations of the involved functions, the general metric-affine theory interpolates continuously between the metric ($K=1$) and the Palatini ($K = 1-(3/2)f'^2/f$) theories. In accordance with Ref.~\cite{Langvik:2020nrs}, we find that these scenarios correspond to
\begin{subequations}
\begin{align} 
\label{eq:g_metric}
    \tilde{f} =& c f^2 - \frac{1}{4c}\,, \qquad \text{(metric)} \\  
\label{eq:g_Palatini}
    \tilde{f} =& c f\,,\, \qquad \qquad \text{(Palatini)} 
\end{align}
\end{subequations}
with $c$ a constant. As an important case, the metric formulation can be obtained in the limit in which the constant part of $\tilde{f}$ is large. More specifically, without loss of generality we can consider $\tilde{f}(h) = \beta/\kappa + \tilde{f}_1(h)$, where $\tilde{f}_1$, $f$ are arbitrary functions of $h$. If $|\beta|/\kappa \gg \tilde{f}_1, f$, then 
\be
\label{eq:K_limit}
    K = 1 
    - \frac{3\kappa}{\beta} \, f' \tilde{f}' 
    + \mathcal{O}\left(\frac{\kappa}{\beta}\right)^{2} \, ,
\ee
and thus the metric-affine theory approaches the purely metric theory when $|\beta| \to \infty$. With $f$, $\tilde{f}$ given by \eqref{def:fg}, the Palatini formulation corresponds to $\beta = \tilde{\xi}/\xi$, of which $\beta=\tilde{\xi}=0$ is only a special case. Consequently, as $\beta$ ranges from $-\infty \to \tilde{\xi}/\xi \to \infty$, the metric formulation is continuously deformed to the Palatini one and back.

\vspace{-0.1cm}
\section{Corrections to vacuum decay in metric-affine gravity}
\label{sec_bounce}
\vspace{-0.1cm}

\begin{figure*}[ht!]  
\begin{center}
\includegraphics[scale=0.4765]{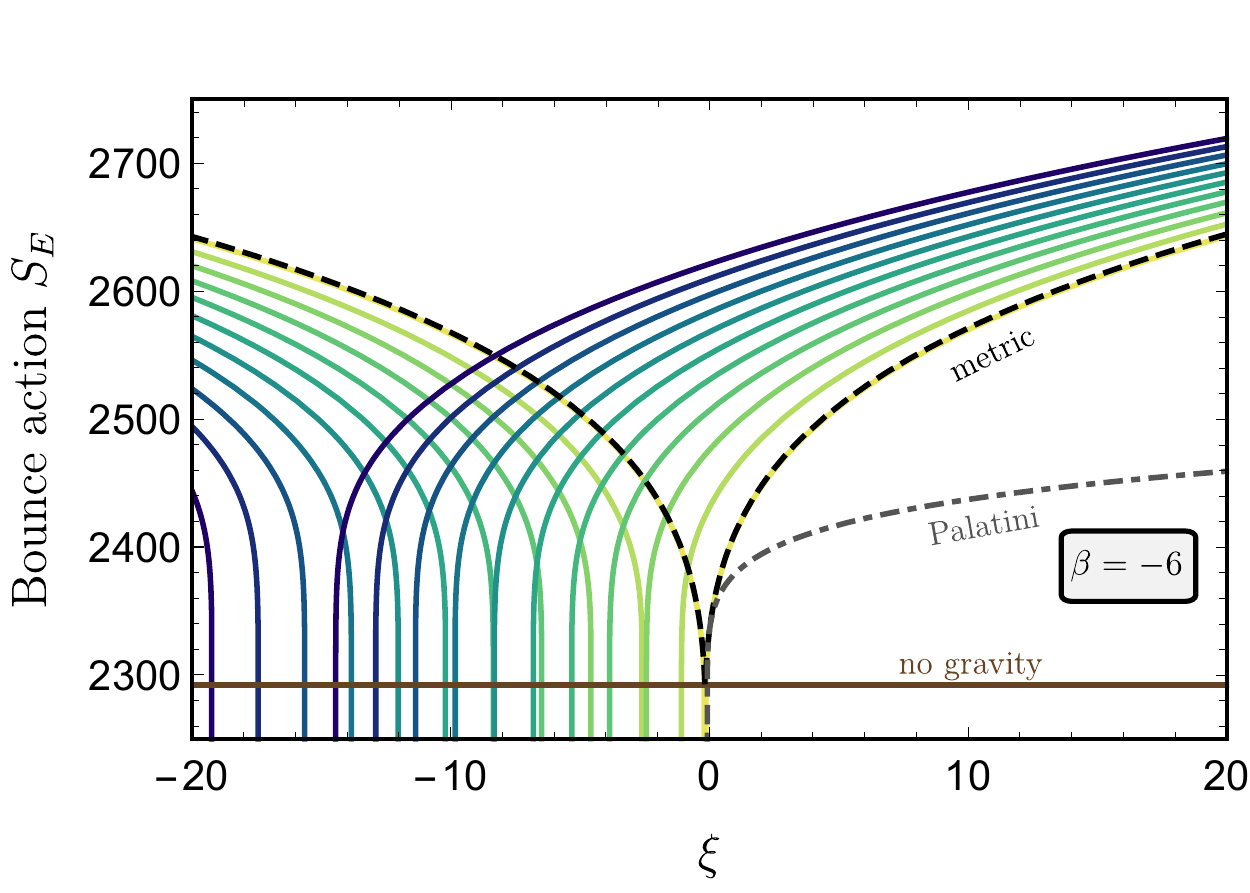}
\hspace{0cm}\includegraphics[scale=0.4535]{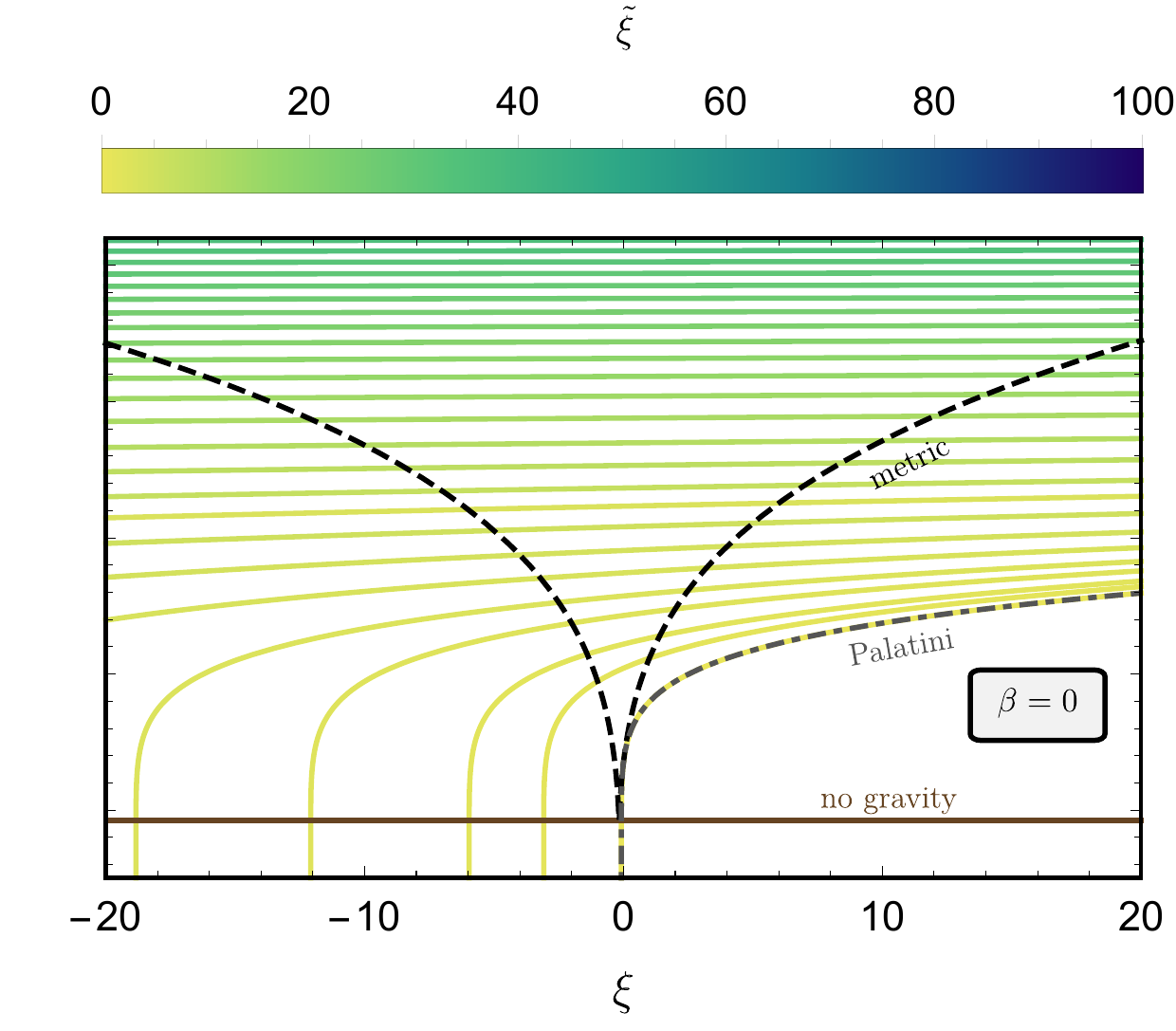}
\hspace{-0.1cm}
\includegraphics[scale=0.472]{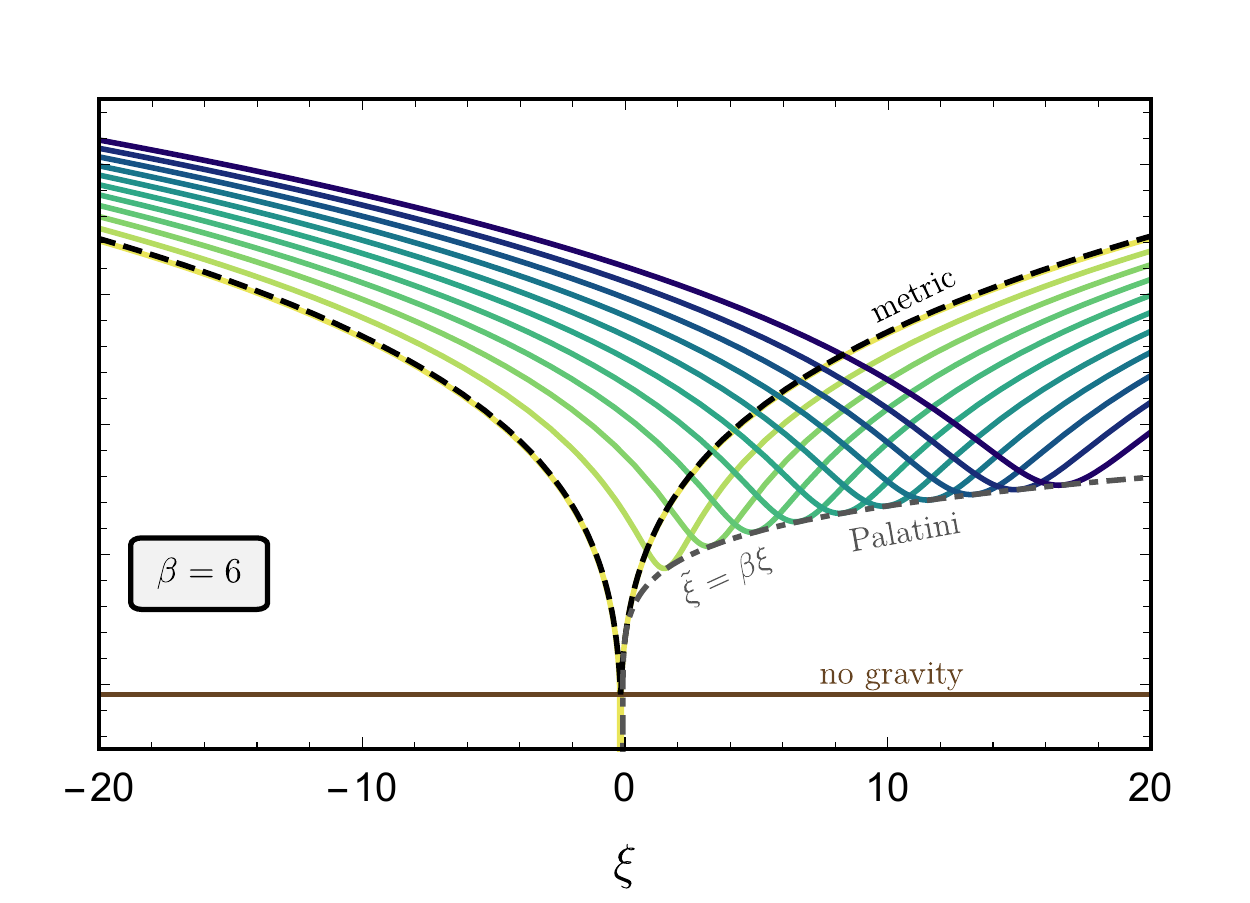}
\caption{The minimal bounce action~\eqref{eq:S_E_exp} for $\b =0, \pm 6$ and various values of $\xi$ and $\tilde{\xi}$. The horizontal line in brown indicates the bounce action in the absence of gravity, $S_0 = 8\pi^2/(3|\l(\mu)|)$. The metric and Palatini cases are represented by the black-dashed and gray-dot-dashed curves, respectively. }
\label{fig:1}
\end{center}
\end{figure*}

To compute the minimal bounce action, we will look for $O(4)$ symmetric solutions~\cite{Coleman:1977th}, with the line element
$\td s^2 =\td r^2 + \rho^2 \td \Omega_3^2$, where $\td\Omega_3^2$ denotes the line element of the unit 3-sphere and the Higgs field depends only on the radial coordinate $r$, \textit{i.e.} $h=h(r)$. In this background, the metric Ricci scalar reads $R = 6 (1-\r \r''-\r'^2)/\r^2$ and the Euclidean action can be recast as
\be
\label{eq:Action_4}
    S_E 
    = 2\pi^2 \int \td r \,\r^3 \Bigg[ 3f \, \frac{\r \r'' + \r'^2 -1}{\r^2} + \frac{1}{2} K h'^2 + V \Bigg]\,.
\ee
The bounce solution is determined by the following equations of motion\footnote{The $\r$ equations of motion follow from the $rr$ component of the Einstein equations, but can also be derived by varying the action~\eqref{eq:Action_4}.}
\begin{subequations}
\label{eq:EoM}
\begin{align}
\label{eq:EoM_r}
    \r'^2 &= 1 +\frac{ \r^2}{3f} \left(\frac{1}{2}K h'^2 - V - 3\frac{\r'}{\r} f' h' \right)\,, \\
\label{eq:EoM_h}
    h'' &= - 3 \frac{\r'}{\r} h' + \frac{1}{K}\left(-\frac{1}{2}K'h'^2 + V' - \frac{R}{2} f' \right)\,.
\end{align}
\end{subequations}
To obtain the bounce action, we will adopt the perturbative method proposed in Ref.~\cite{Isidori:2007vm} and look for solutions as a series in $\k$, \textit{i.e.}
\begin{subequations}
\begin{align}
\label{eq:approx_h}
    h(r) =& h_0(r) +\k h_1(r) + \mathcal{O}(\k^2)\,, 
    \qquad \\ 
\label{eq:approx_r}
    \r(r) =& r +\k \r_1(r)  +\k^2 \r_2(r) + \mathcal{O}(\k^3)\,,
\end{align}
\end{subequations}
This approach is suitable when the gravitational corrections are relatively small, and, as we will demonstrate, this technique is adequate for elucidating the differences that emerge due to the inclusion of the Holst invariant. In a similar vein, the bounce action~\eqref{eq:Action_4} can be expanded as
\be
\label{eq:S_E_exp}
    S_E = S_0 + \kappa S_1 + \mathcal{O}(\k^2)\,.
\ee

The leading order solution\footnote{See also~\cite{Tetradis:2023fnu} for exact solutions for vacuum decay in Higgs-like unbounded potentials.} $h_0(r)$ of~\eqref{eq:EoM} is the so-called Fubini instanton~\cite{Fubini:1976jm, Lee:1985uv} 
\be
\label{eq:fubini}
    h_0(r) = \sqrt{\frac{2}{|\l|}}\frac{2\mu}{1 + \mu^{2}r^2}\,,
\ee
with $\mu$ being an arbitrary scale of the bounce. It solves the equation of motion in the absence of gravity ($\rho_0 = r$) and with $V(h) = \l h^4/4$ assuming that the Higgs quartic coupling $\l$ is constant and negative. The leading order contribution to the action~\eqref{eq:Action_4} is
\be
\label{eq:Action_S_0}
    S_0 = 2\pi^2 \int {\td} r\, r^3 \left(\frac{h_0'^2}{2}+V(h_0) \right) = \frac{8\pi^2}{3 |\l|}\,,
\ee
and gives the bounce action in the absence of gravity. To obtain the gravitationally corrected action one must account for the running of $\l$~\cite{Salvio:2016mvj,Rajantie:2016hkj}. We will evaluate $\l$ at the scale of the bounce $\mu$ and then minimize the action with respect to $\mu$. The running of $\l$ is computed at a 3-loop level~\cite{Buttazzo:2013uya} with the relevant parameters taken from~\cite{ParticleDataGroup:2022pth}.

Evaluating the gravitational correction $S_1$ relies on the specific form of $f$ and $\tilde{f}$, for which we will assume the form~\eqref{def:fg} when needed. We will assume that the leading order gravitational corrections to the kinetic function~\eqref{eq:K} can be expressed as
\be
\label{eq:K_ser}
    K = 1 +\k K_1 h^2 + \mathcal{O}(\k^2)\,,
\ee
where $K_1$ is a dimensionless constant. Indeed, with $f$ and $\tilde{f}$ given by~\eqref{def:fg}, we have that
\be
    K_1 \equiv -6\frac{\xi^2 + 2 \beta  \xi  \tilde{\xi} - \tilde{\xi}^2}{1 + \beta ^2}\,.
\ee
Note that for large $\beta$ we recover Eq.~\eqref{eq:K_limit}. At order $\mathcal{O}(\kappa)$, the equation of motion~\eqref{eq:EoM_r} is
\be
    \rho_1' = \frac{1}{6} r^2 \left( \frac{1}{2} h_0'^2 - V(h_0) - 3 f'(h_0)h_0'\right)\,,
\ee
independently of the shape of $\tilde{f}$.  For the Fubini bounce~\eqref{eq:fubini}, it is solved by
\be
\label{eq:rho_1}
    \rho_1 = \frac{1+6\xi}{3|\lambda|/\m^2}\left(r \frac{\mu^2 r^2 - 1}{( \mu^2 r^2+1)^2} + \mu^{-1} {\rm \arctan}(\mu r)\right)\,.
\ee
Knowing $\rho_1$ is sufficient to compute the $\mathcal{O}(\kappa)$ correction $S_1$ to the action. We checked explicitly that the dependence on $h_1$ can be eliminated by the $h_0$ equations of motion and partial integration. 
This is a general result, however, because $h_0$ minimizes the action in the absence of gravity and thus $S_0[h_0 + \kappa h_1] = S_0[h_0] + \mathcal{O}(\kappa^2)$~\cite{Salvio:2016mvj}. In all, we obtain that
\bea
\label{eq:result}
    S_{1} 
    &= \frac{32 \p^2 \mu^2 }{45 \l^2(\mu)} \left( (1+6\xi)^2 + 6 K_1 \right)\, 
    \\
    &=  \frac{32 \p^2 \mu^2 }{45 \l^2(\mu)} \left( (1+6\xi)^2
    - 36\frac{\xi^2 + 2 \beta  \xi  \tilde{\xi}- \tilde{\xi}^2}{1 + \beta ^2} \right)\,,
\eea
where $K_1$ encodes the modifications resulting from an independent connection, \textit{i.e.}, setting $K_1 = 0$ recovers the metric case.

The gravitational correction \eqref{eq:result} can be negative for certain values of model parameters. If this happens, then the action cannot be minimized with respect to $\mu$ and the adopted perturbative approach is not applicable. However, by minimizing $S_1$ with respect to $\xi$, it is straightforward to show that  $S_{1}$ is always positive when
\be
    \beta \tilde{\xi} \geq 1/12 \,.
\ee
Otherwise, the positivity of the gravitational correction $S_1$ can be achieved only in certain regions of the parameter space. Two special cases warrant being considered more closely:
\begin{enumerate}
\item $\tilde{\xi}=0$: A minimally coupled Holst term $\tilde{f}(h)=\beta/\k$ is probably the simplest scenario. The region allowed by the positivity of $S_1$ is
\be
\label{eq:xi_ineq_j1}
    \left|\xi + \frac{1}{6} + \frac{1}{6\b^2}\right| \geq \frac{\sqrt{1+\b^2}}{6\b^2}\,.
\ee
For $\b\ll 1$, this gives $\xi \leq -1/(3\b^2)$ or $\xi \geq -1/12$, while, when $\b\gg 1$, only a narrow region around the conformal coupling $\xi = -1/6$ is forbidden, that is, $|\xi + 1/6| \geq 1/(6\b)$.

\item $\b=0$: In this case, the contribution from the Holst term
\be
\label{eq:result_2}
S_1 =  \frac{32 \p^2 \mu^2}{45 \l^2(\mu)}\left( 1+12\xi +36\tilde{\xi}^2 \right) \,,
\ee
is always positive and will thus always improve the stability of the SM vacuum when compared to the Palatini case.  This special case is depicted in the middle panel of Fig.~\ref{fig:1}
However, as in the Palatini limit, the positivity of $S_1$ implies a strict lower bound
\be
\label{eq:xi_ineq_b}
    \xi \geq -1/12-3\tilde{\xi}^2\,.
\ee
\
\end{enumerate}

The minimal bounce action is shown in Fig.~\ref{fig:1} with the running of $\lambda$ computed at the 3-loop level~\cite{Buttazzo:2013uya}. It shows that the metric limit $\beta \to \infty$ is realized quite well already for $|\beta| = 6$ when $\tilde \xi \approx 0$. For $\tilde \xi \gg 1$, we see that the bounce action is typically enhanced when $\b$ and $\xi$ have the opposite signs,  thus improving the stability of the vacuum. Additionally, in comparison to the metric case, the stability is improved when $|\xi|<|\tilde{\xi}|$. In all depicted cases, the regions in which $S_1 < 0$ can be observed: When $\b = 6$, this region exists only for the $\tilde\xi = 0$ line and is contained in a narrow range around $\xi = -37/216$. In the $\b = -6$ case, a parameter region with $S_1 < 0$ can be observed for every $\tilde\xi$. The disallowed $\xi$ range varies with $\tilde\xi$. Since the theory is independent of the sign of $\tilde{f}$, then the $\b = 6$ panel covers the $\b = -6$ case with $\tilde \xi \in [-100,0]$ and vice versa. 

Finally, it is important to point out that, as in the action \eqref{eq:Action_3b}, the contributions from mass dimension 2 terms constructed from torsion and non-metricity that can appear in metric-affine gravity can be reduced to a non-canonical kinetic term in a metric theory~\cite{Rigouzzo:2022yan}. This implies that our results can be straightforwardly extended to include corrections to vacuum stability from such terms by computing their contribution to the small $\k$ expansion \eqref{eq:K_ser}.

\vspace{-0.1cm}
\section{Conclusions}
\label{Conclusions}
\vspace{-0.1cm}

We analyzed the stability of the electroweak vacuum in metric-affine gravity, where the Higgs boson is expected to have an additional non-minimal coupling to the Holst invariant. This scenario can be reformulated in terms of an equivalent metric theory with a non-canonical kinetic term, where the gravitational corrections to the bounce action can be studied with established perturbative methods. Our results show that the stability of the electroweak vacuum in metric-affine gravity is improved across a wide range of model parameters.

A non-minimally coupled Holst term provides a class of models that continuously connects metric and Palatini gravity. We find that the limiting case of Palatini gravity displays the mildest improvement to vacuum stability.

\vspace{-0.1cm}
\section*{Acknowledgments} 
\vspace{-0.1cm} 
\noindent   We thank Tomi Koivisto for useful comments. This work was supported by the Estonian Research Council grants SJD18 and PSG869.

\bibliography{refs_vac_dec}{}

\end{document}